\documentclass[a4paper]{article}
\usepackage[utf8]{inputenc}
\usepackage[T1]{fontenc}

\usepackage{graphicx}

\usepackage{subfig}
\usepackage{wrapfig}

\usepackage{hyperref}

\usepackage{color}
\usepackage{listings}
\definecolor{mygrey}{rgb}{0.95,0.95,0.95}
\definecolor{myorange}{rgb}{0.9,0.1,0.1}
\definecolor{mygreen}{rgb}{0.1,0.4,0.1}
\usepackage{authblk}

\begin{document}

\title{Software-Distributed Shared Memory for Heterogeneous Machines: Design and Use Considerations}

\author[1,2]{Loïc Cudennec}

\affil[1]{CEA, LIST\protect\\F-91191, PC 172, Gif-sur-Yvette, France}
\affil[2]{DGA MI, Department of Artificial Intelligence\protect\\BP 7, 35998 Rennes Armées, France\protect\\\url{loic.cudennec@def.gouv.fr}}

\date{\today}

\maketitle

\begin{abstract}
Distributed shared memory (DSM) allows to implement and deploy
applications onto distributed architectures using the convenient
shared memory programming model in which a set of tasks are able to
allocate and access data despite their remote localization. With the
development of distributed heterogeneous architectures in both HPC and
embedded contexts, there is a renewal of interest for systems such as
DSM that ease the programmability of complex hardware. In this report,
some design considerations are given to build a complete software-DSM
(S-DSM). This S-DSM called SAT (Share Among Things) is developed at
CEA (the French Alternative Energies and Atomic Energy Commission)
within the framework of European project M2DC (Modular Microserver
DataCentre) to tackle the problem of managing shared data over
microserver architectures. The S-DSM features the automatic
decomposition of large data into atomic pieces called chunks, the
possibility to deploy multiple coherence protocols to manage different
chunks, an hybrid programming model based on event programming and a
micro-sleep mechanism to decrease the energy consumption on message
reception.
\end{abstract}

\section{Introduction}

Shared memory is a convenient programming model in which a set of
tasks (processes, threads) are able to access (allocate, read, write)
a common memory space. This is quite straightforward in the classic
Von-Neumann architecture in which physical memories are shared among
the processing units. However, when coping with distributed
architectures, processing units can not directly access remote
memories using a local address space. Some intermediate hardware or
software systems are needed to transparently manage access
requests. Such systems include Software-Distributed Shared Memory
(S-DSM) as proposed in the late eighties with
IVY~\cite{DBLP:conf/icpp/Li88} and more recently studied for modern
architectures with Grappa~\cite{DBLP:conf/usenix/NelsonHMBCKO15} and
Argo~\cite{DBLP:conf/hpdc/KaxirasKNRS15}.

S-DSM can be seen as a distributed middleware application standing
between the user code and the diversity of OS primitives and libraries
that manage local memory, remote services (eg. RDMA, one-sided
communications) and message passing (eg. MPI). It provides a
platform-agnostic global, logical address space to the user code,
which constitutes a step towards single-system image and operating
systems (SSI) originally introduced for computing clusters two decades
ago~\cite{DBLP:journals/ijhpca/BuyyaCJ01}. The abstraction of the
platform not only aggregates remote memories into a virtual space, it
also tackles the problem of heterogeneity at different levels, from
the communication medium to the data representation. This simplifies
the management of data in distributed applications and the possibility
to transparently reuse code and deploy on different heterogeneous
platforms.

S-DSM runtimes usually introduce significant overheads mainly due to
the increase of the number of coherence protocol messages compared to
message-passing (MP) applications. With the recent development of
high-speed network and new efficient protocols to access remote
memories, modern S-DSM are now able to match or exceed the performance
of MP-designed applications. In this work~\cite{CudennecParaMo2020},
the S-DSM proposed in this report is evaluated together with a
classical Open MPI implementation and ZeroMQ, a lightweight MP
implementation originally designed for embedded systems. Results show
that this S-DSM performs better than the Open MPI implementation and
competes with the ZeroMQ implementation. In this report we give some
information on how this S-DSM is designed and how applications are
implemented over the API.

This report is organized as follows: section~\ref{sec:model} presents
the S-DSM programming model as well as some elements on the way it is
implemented in the runtime. Section~\ref{sec:application} gives some
information on how to write an application on top of the
S-DSM. Finally section~\ref{sec:conclusion} concludes this report and
gives some insights.

\section{S-DSM programming model}
\label{sec:model}

\begin{figure*}
\centering
  \includegraphics[width=0.5\textwidth]{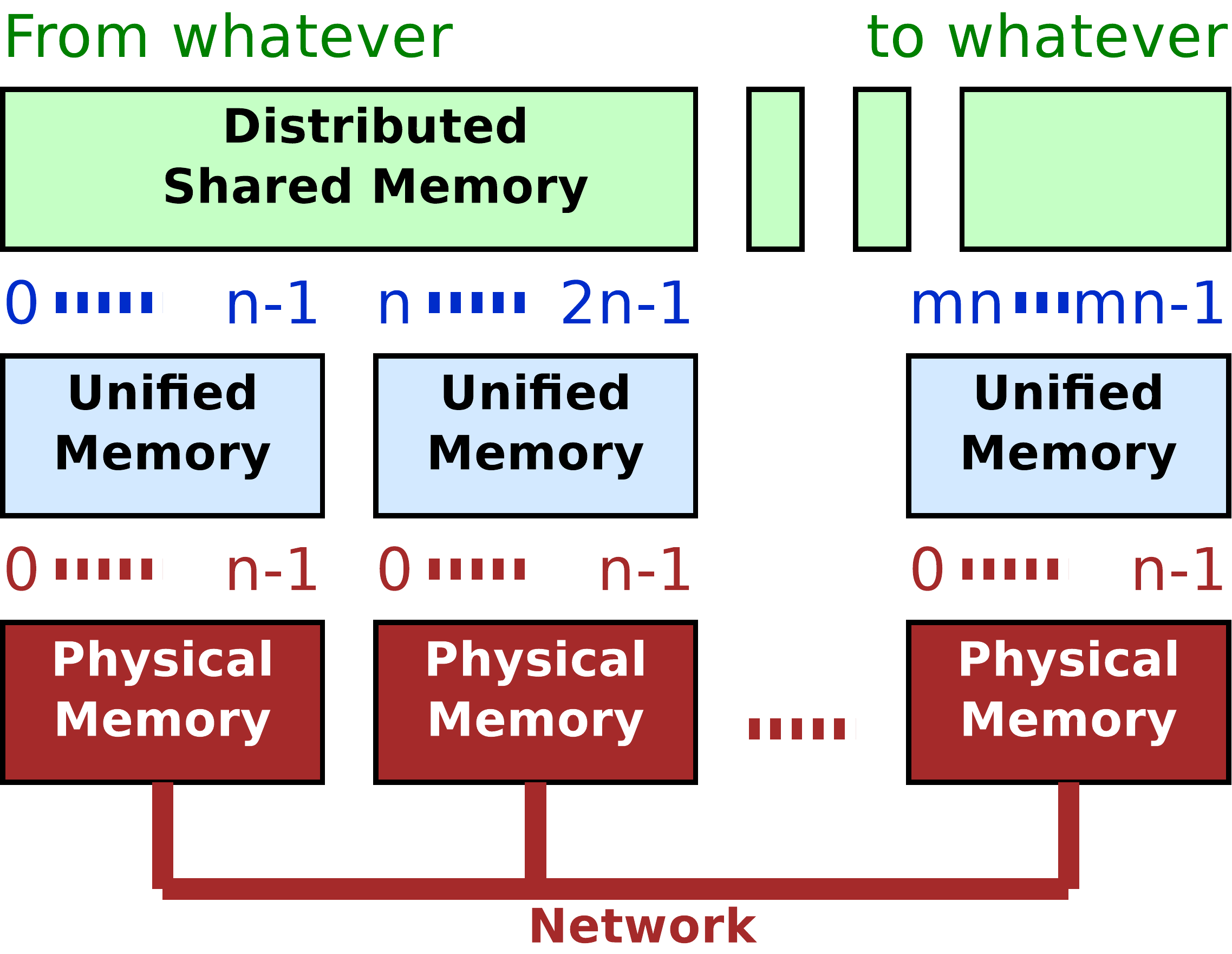}
  \caption{S-DSM as a middleware to unify remote memories.}
  \label{fig:sdsm_concept}
\end{figure*}

In this work we propose a S-DSM~\cite{DBLP:conf/europar/Cudennec17}
designed to ease the programmability of distributed heterogeneous
platforms. Figure~\ref{fig:sdsm_concept} represents a distributed
architecture in which physical memories are connected via a
network. Each memory has a specific address space. A first abstraction
layer concatenates the memory address spaces into a global one. The
resulting address space is however hardware-dependent: the user has to
cope with NUMA (Non-Uniform Memory Access) and manage data locality
and replication. The S-DSM abstraction layer builds a logical address
space, which is not dependent from the underlying hardware. This
system hides the data localization, replication and transfer and
provides a simple Posix-like interface.

The implementation of the proposed S-DSM stands at the user level, as
a portable way to deploy onto different operating systems. A set of
clients running the application code are connected to a peer-to-peer
(P2P) network of data management servers. Shared data are decomposed
into chunks of any size to allow the allocation of large memory
segments and the limitation of false-sharing. Chunks are independently
managed under the supervision of a consistency protocol. It is
possible to deploy several consistency protocols for the same
application to manage different chunks. The default protocol is a
$4$-state home-based protocol. These concepts have been used in
several data-sharing systems, including
OceanStore~\cite{DBLP:conf/asplos/KubiatowiczBCCEGGRWWWZ00},
DSM-PM2~\cite{DBLP:conf/ipps/AntoniuB01} and
JuxMem~\cite{DBLP:journals/scpe/AntoniuBJ05}.

The API is based on the scope consistency
model~\cite{DBLP:conf/spaa/IftodeSL96} in which accesses to shared
data are protected within an acquire-release scope. The S-DSM malloc
primitive can be called from every client, taking into parameter an
address in the global logical space. It is also possible to use a
built-in symbolic table to identify shared data using plain text
instead of the logical address space. Distributed synchronization
primitives are provided such as rendez-vous, barriers and signals. The
signal mechanism has also been used to implement the publish-subscribe
model applied to chunks~\cite{DBLP:conf/europar/Cudennec18}: each time
a chunk is modified, a notification is sent to the subscribers. This
allows the design of applications based on a mix of the shared memory
and the event-based programming models. The S-DSM has been implemented
over the MPI runtime to manage communications between nodes. It
inherits from the task model, the automatic deployment and
bootstrapping of the communication world, the management of multiple
message queues and the optimizations for message delivery. It also
provides MPI tools for debugging low-level communications and the MPI
runtime conveniently redirect standard outputs of each remote process
to the master node. On top of that, S-DSM events are logged, processed
and can be used to debug and optimize applications.

This S-DSM has been used to implement a dataflow-oriented video
processing application. Communication channels are instantiated as
shared data. Tasks that write to an output channel use the regular
shared memory access primitive while tasks that read from an input
channel rely on the publish-subscribe mechanism and get notified each
time the channel has been modified to retrieve a new token. This
application has been used to evaluate the performance of the S-DSM,
experiment, and build a demonstrator for scientific and industrial
forums.

This work takes place within the context of the European project
M2DC~\cite{DBLP:journals/mam/OleksiakKPABBFP17} in which a microserver
architecture such as the Christmann RECS$\mid$Box Antares Microserver
and a software stack is proposed. A microserver is composed by a
rackable backplane (1U/4U) providing power supply and networking
capabilities to a set of slots that can host heterogeneous expansion
cards such as high performance CPU, low-power CPU, manycore
processors, GPGPU or FPGA. In these systems, there is no central
physical memory and the application has to manually cope with the
management of data. Therefore, middlewares such as S-DSM provide a
convenient abstraction layer to unify the memories of the different
expansion slots. Hybrid programming is still required to get access to
the accelerators: the S-DSM is able to manage data between nodes with
a CPU. However, moving data from a CPU host and an accelerator (GPGPU,
FPGA) is still the responsibility of the user, as it is commonly done
for MPI/OpenMP or MPI/CUDA in HPC systems. In this
work~\cite{LenormandRSP2020}, a system is proposed to transparently
manage these interactions between hosts and FPGAs based on this S-DSM,
providing a full DSM implementation among the processing elements,
however this is not in the scope of this report.

Security and access control to the S-DSM is studied
in~\cite{DBLP:conf/icete/StanCS20}, providing a transparent layer
implementing attribute-based encryption (ABE) between the S-DSM API
and the user code. More information can be found in the paper.

In the following sections, we describe some parts of the S-DSM that
has been developed to study data management strategies over
distributed heterogeneous platforms.

\subsection{General S-DSM layout}

\begin{figure*}
\centering
  \includegraphics[width=0.5\textwidth]{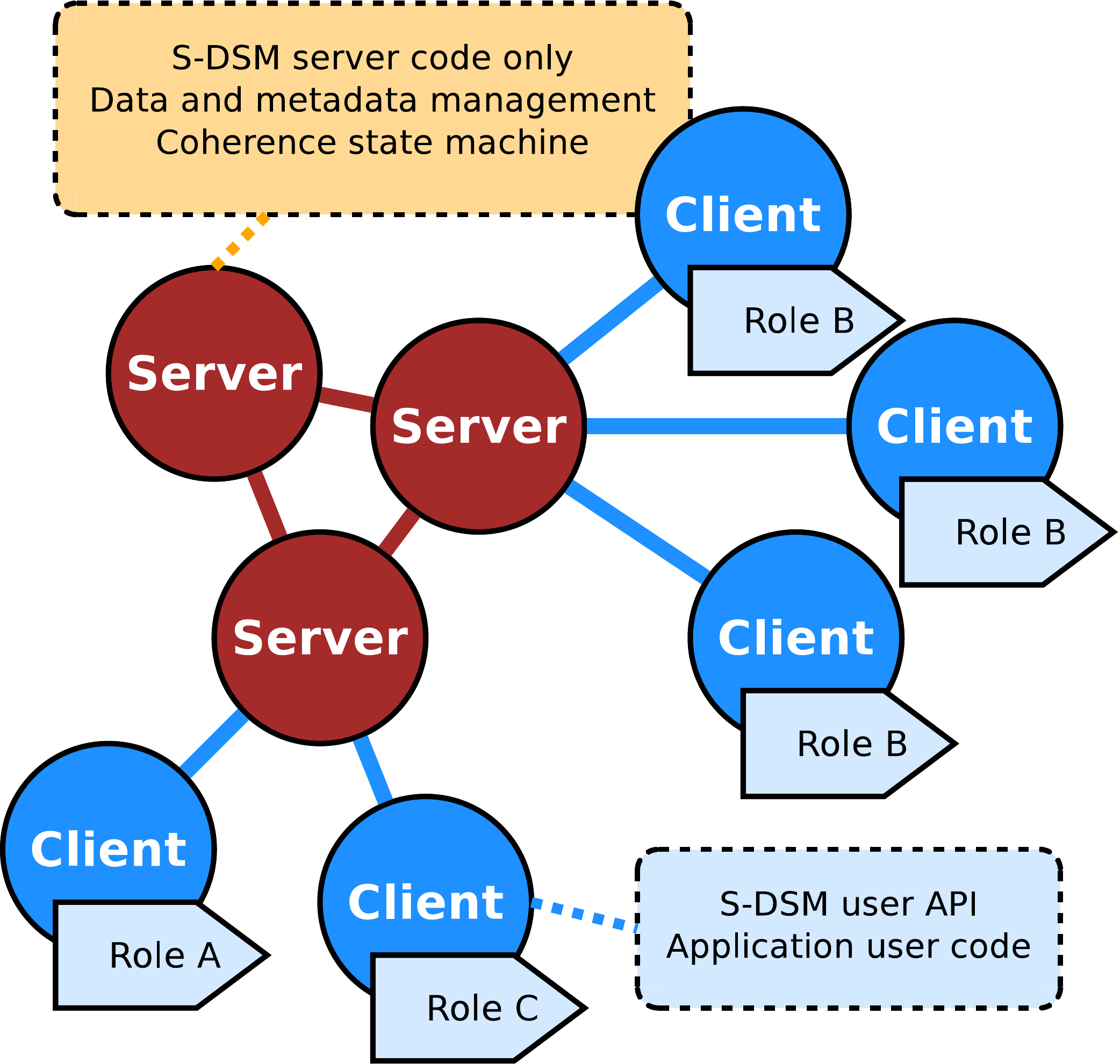}
  \caption{S-DSM super-peer topology.}
  \label{fig:sdsm_super_peer_topology}
\end{figure*}

The S-DSM developed at CEA has been designed as an experimental
platform for shared data management in emerging distributed
heterogeneous architectures. It is possible to implement different
consistency protocols and deploy several of them during the same run
to manage different shared data. It is implemented at the user-level
and does not require any modification of the OS kernels or system
libraries to remain portable and easy to deploy on multiple systems
found in heterogeneous platforms.

The S-DSM is organized as a semi-structured super-peer topology made
of a peer-to-peer network of servers and a set of clients, as
presented in Figure~\ref{fig:sdsm_super_peer_topology}. Clients run
the user code and provide the interface to the shared memory. Servers
execute coherence automata and manage data and metadata.

Applications written for the S-DSM are based on a simple task model,
each task instance running as a S-DSM client. Parallelism comes from
the multiplicity of instances for each task. Prior to a deployment, a
logical topology has to be defined thanks to an instantiating step and
a mapping step. The performance of the application largely depends on
these two steps. Instantiating the application consists in choosing
the number of instances to be created for each task, including the
number of instances of S-DSM metadata servers. These instances have to
be connected together, each instance of a client connected to a S-DSM
server instance, in order to form a logical topology. This topology is
thereafter mapped onto the physical resources and then effectively
deployed. These deployment steps are complex even when coping with
rather small applications and small heterogeneous clusters. In this
paper~\cite{DBLP:conf/europar/TrabelsiCB19}, a compilation toolchain
is proposed to explore and optimize the deployment of the S-DSM onto
distributed heterogeneous resources, using local search
algorithm. This results in a Pareto front from which it is possible to
pick a solution that fits to specific computing performance and energy
consumption constraints.

This implementation of the S-DSM is not thread-safe, meaning that
threads belonging to the same task cannot concurrently access the
shared memory without explicit synchronization. If such limitation
becomes critical for an application, concurrent threads can be placed
into different S-DSM clients and co-located onto the same physical
resource to benefit from locality.

The S-DSM is designed as an application helper or a temporary service:
it starts when a call to the bootstrapping function is made and
terminates when all clients have notified a termination.

\subsection{Atomic slices of memory: chunks}

\begin{figure*}
\centering
  \includegraphics[width=1\textwidth]{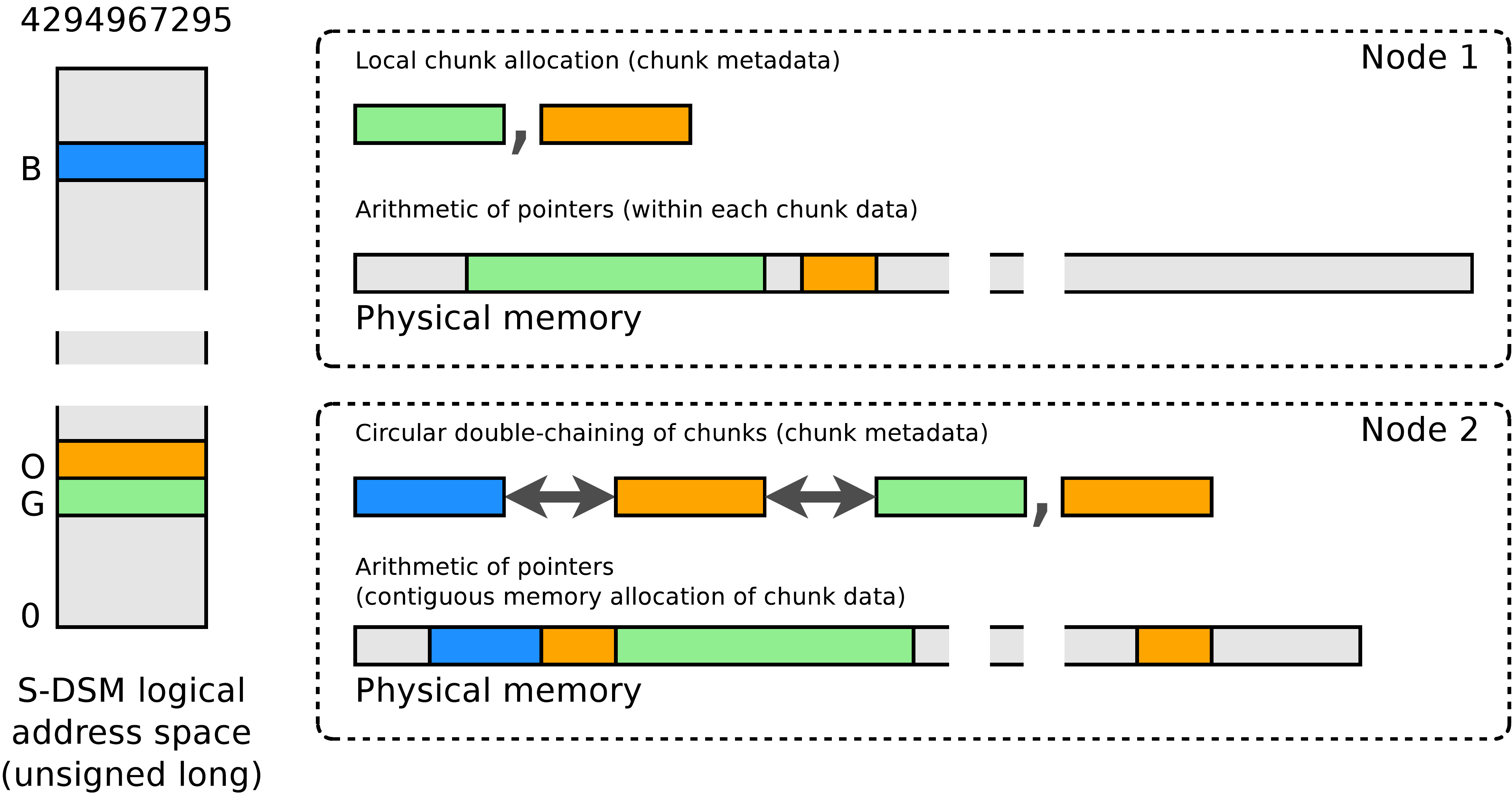}
  \caption{From the S-DSM logical address space to the local memory.}
  \label{fig:sdsm_logical_address_space}
\end{figure*}

Shared memory, as a global space containing data, is divided into
atomic pieces called \emph{chunks}. Each chunk is identified by an
address in the S-DSM logical address space. In this implementation,
the logical address space contains all possible values of an unsigned
long, as defined by ANSI C on modern architectures. Chunks can be of
any size and any type, as a multiple of a C~\emph{char}
type. Figure~\ref{fig:sdsm_logical_address_space} represents the S-DSM
logical space and how chunks can be differently mapped into local
memories. Chunks $G$ and~$O$ are contiguously allocated in the S-DSM
logical space while $B$ is not. It is possible to do arithmetic onto
the S-DSM logical space, for example $@O=@G+1$. However, this logical
mapping is totally independent from the local mapping on each node.

On $Node1$, chunks $G$ and~$O$ have been independently allocated and
their data are mapped in arbitrary non-contiguous addresses of the
local memory. On $Node2$, chunks $B$, $O$ and~$G$ have been allocated
as a chunk~\emph{chain}. A chunk chain is a sequence of chunks that
ensures a contiguous allocation of data in memory. In practice, it is
a circular double-chain of chunks. In this configuration, it is
possible to do arithmetic of pointers from the data pointed by chunk
$B$ directly followed by chunks $O$ and~$G$. Chunk chains can be
allocated on user nodes to locally merge data or build complex
patterns. However they do not exist outside the node as chunks are
independently managed by the S-DSM. It is also possible to allocate a
chunk several times into the local memory, even within different chunk
chains. In that case, chunk data will be map in different places in
the local memory and kept consistent (a write to an instance of the
chunk will also be applied to the other instances of the chunk on the
local node).

\begin{figure*}
  \centering
  \begin{lstlisting}
    Consistency_t * consistency = newHomeBaseMESI();
    Chunk_t * chunk = NULL;

    unsigned long chunkid = 42;
    size_t size = sizeof(char) * SIZE;
    chunk = MALLOC(consistency, chunkid, size);

    UnsignedLongList_t * idlst = {16, 81, 56878};
    UnsignedLongList_t * sizelst = {24, 91, 54};
    chunk = MALLOC_LST(consistency, idlst, sizelst);

    unsigned int nbchunks = 1;
    chunk = LOOKUP(consistency, chunkid, nbchunks);

    chunk = LOOKUP_LST(consistency, idlst);
    
  \end{lstlisting}
  \caption{Malloc and lookup.}
  \label{fig:sdsm_malloc_listing}
\end{figure*}

Several functions are provided by the client interface of the S-DSM to
allocate chunks. The main functions given in
Figure~\ref{fig:sdsm_malloc_listing} return a pointer to a chunk that
can be later used to access data.

\begin{description}

\item[MALLOC] allocates chunks starting at address \verb!@baseid! in
  the S-DSM logical memory space for a total size of data given by the
  \verb!size! parameter. The number of contiguous chunks allocated is
  calculated based on the default chunk size set in the S-DSM and the
  last chunk size is appropriately calculated so that no memory space
  is wasted. Allocated chunks are linked within a chunk chain. If the
  exact same chunk chain has already been \emph{locally} allocated, it
  does not allocate new chunks: it returns the corresponding chunk
  chain. If other local chunk chains contain one or more chunks from
  this chunk chain, chunks are replicated in local memory so that the
  data of the chunk chain is contiguous in local memory.

\item[MALLOC\_LST] allocates chunks using the list of addresses
  \verb!idlst!. The size of individual chunks is given by the
  \verb!sizelst!  parameter. In this example, chunk \verb!@81! is
  allocated with data size $91$. The allocator does a round-robin onto
  list \verb!sizelst! if it is smaller than \verb!idlst!. Allocated
  chunks are linked within a chunk chain and properties are the same
  than for MALLOC.

\item[LOOKUP] returns previously allocated chunks in the S-DSM (chunks
  that have been allocated and written by any process). Compared to
  the MALLOC primitive, LOOKUP does not require to specify the size of
  the data. This is convenient if it is calculated at runtime and not
  known by a process. If more than one chunk is requested ($nbchunks >
  1$) a chunk chain is returned made of contiguous addresses in the
  S-DSM memory space and starting at address \verb!@chunkid!.

\item[LOOKUP\_LST] is similar to LOOKUP except that it takes a list
    of chunk addresses instead of a base address.

\end{description}

A consistency protocol must be set to allocate chunks. It can be a
different protocol for each chunk, hence implementing a
multi-consistency system. This consistency protocol will be used for
each access to the chunk and will drive the S-DSM behaviour regarding
the data and metadata management.

\subsection{Access modes}

A chunk is a structure that hosts consistency state information about
shared data and a local pointer to the data. The user code can access
this pointer following different modes.

\begin{figure*}
  \centering
  \begin{lstlisting}
    chunk = MALLOC(consistency, chunkid, size);
    
    WRITE(chunk);

    /** chunk->data exists
    *   but values may not be initialized
    **/
    
    for (i = 0; i < N; i++) {
      chunk->data[i] = i;
    }
    RELEASE(chunk);

    /** do not use chunk->data here as
    *   - consistency is not guaranteed
    *   - pointer can be NULL (freed)
    **/
    
    READWRITE(chunk)

    /* chunk->data exists and data are updated */
    
    for (i = 0; i < N; i++) {
      chunk->data[i] = N - chunk->data[i];
    }    
    RELEASE(chunk)

    READ(chunk)

    for (i = 0; i < N; i++) {
      chunk->data[i] = N - chunk->data[i];
    }    

    RELEASE(chunk)

    /** last modification of chunk->data is lost
    *   as it was a read-only scope
    **/
  \end{lstlisting}
  \caption{Scope consistency.}
  \label{fig:sdsm_scope_listing}
\end{figure*}

\begin{figure*}
  \centering
  \begin{lstlisting}
    void * data = calloc(N, sizeof(int));

    MAP(chunk, consistency, baseid,
        N * sizeof(int), data);

    assert(data == chunk->data);
    
    for (i = 0; i < N; i++) {
      data[i] = i;
    }

    PUT(chunk); /* equivalent to WRITE then RELEASE */

    /** data and chunk->data can be used here
    *   however, consistency is not guaranteed
    **/
    
    GET(chunk); /* equivalent to READ then RELEASE */

    /** data might now contain updated values
    *    if other S-DSM processes have modified chunk
    *   consistency is not guaranteed
    **/
  \end{lstlisting}
  \caption{Memory mapping.}
  \label{fig:sdsm_mmap_listing}
\end{figure*}

\begin{figure*}
  \centering
  \begin{lstlisting}
    void * datawrite = calloc(N, sizeof(int));
    void * dataread = NULL;
    size_t size = 0;
    
    initSymbolicTable(sdsm);

    writeSymbol(sdsm, "symbol_name",
                data, N * sizeof(int));

    readSymbol(sdsm, "symbol_name",
              &dataread, &size);
    free(dataread);
    dataread = NULL;
    
    readSymbol(sdsm,
          "uuid-921b4274-84ad-4b04-ac75-f9738da84039",
          &dataread, &size);
  \end{lstlisting}
  \caption{Symbolic table.}
  \label{fig:sdsm_symbol_listing}
\end{figure*}

\begin{description}
  
\item[Scope consistency]~\cite{midway} implies that all accesses must
  be protected between a call to 1) READ, WRITE or READWRITE primitive
  to enter the scope and a call to 2) RELEASE primitive to exit the
  scope. Outside this scope, data consistency is not guaranteed and
  the pointer to the data can be discarded if the S-DSM is running
  short on local memory. In this latter case, the data is present in
  another node of the S-DSM. However, in order to use the data outside
  the scope, a local copy must be made by the user within the
  scope. Examples are given in Figure~\ref{fig:sdsm_scope_listing}.

\item[Memory mapping] is used to keep the data pointer safe outside
  the consistency scope, without data copy (zero-copy). The MAP
  primitive maps the provided data pointer to the chunk chain starting
  by the given chunk. This can be used together with PUT and GET
  primitives that basically behave as WRITE-RELEASE and READ-RELEASE
  empty scopes. Examples are given in
  Figure~\ref{fig:sdsm_mmap_listing}.

\item[Table of symbols] provides an abstraction layer over the S-DSM
  logical address space: shared data are identified by symbols (text,
  as a C string) and a shared built-in table is used to match symbols
  with the corresponding chunks. This symbolic table is stored into
  the S-DSM as a regular shared data. It is close to the memory
  mapping access model in which the data pointer is
  preserved. Examples are given in
  Figure~\ref{fig:sdsm_symbol_listing}.

\end{description}

All accesses to shared data in the S-DSM are achieved under the
supervision of the consistency protocol that has been set when
allocating the data. The default protocol is a home-based MESI
protocol~\cite{DBLP:books/daglib/0091765}. This protocol allows multiple
parallel reads and an exclusive single write, hence implemented with
the four Modified, Exclusive, Shared, Invalid states. Each shared data
is managed by a specific node in the system also referred as the
home-node. In our implementation this node is calculated as a modulo
on the S-DSM server list.

\subsection{Synchronization objects}

\begin{figure*}
  \centering
  \begin{lstlisting}
    void A () {
      WRITE(chunk);
      modify(chunk->data);
      RELEASE(chunk);
      wakeup(RDV_ID);
      barrier(BAR_ID, clientGetClientNr());
    }
  \end{lstlisting}
  \begin{lstlisting}
    void B () {
      rendezvous(RDV_ID);
      READWRITE(chunk);
      display(chunk->data);
      modify(chunk->data);
      RELEASE(chunk);
      barrier(BAR_ID, clientGetClientNr());
    }
  \end{lstlisting}
  \begin{lstlisting}
    void C () {
      barrier(BAR_ID, clientGetClientNr());
      READ(chunk);
      display(chunk->data);
      RELEASE(chunk);
    }
  \end{lstlisting}

  \caption{Synchronization of $3$~tasks using a rendezvous and a
    barrier.}
  \label{fig:sdsm_synchronization_listing}
\end{figure*}

Processes can synchronize in the S-DSM using distributed
synchronization objects. The two main objects are rendezvous and
barriers, implemented using Raynal's distributed
algorithms~\cite{DBLP:books/daglib/0032304}.

\begin{description}

\item[sleep] is used to synchronize on a given rendezvous. Rendezvous
  are identified by an unsigned int. The process hangs until a wake up
  signal is received.

\item[wakeup] is used to wake up all processes that are sleeping on a
  given rendezvous.

\item[barrier] is used to synchronize a given number of
  processes. Processes that enter a barrier hang until the number of
  processes expected in the barrier is reached. Barriers are
  identified by an unsigned int, in a different identifier space than
  rendezvous.

\end{description}

Figure~\ref{fig:sdsm_synchronization_listing} gives an example of
explicit synchronization between three tasks. Task~$B$ waits for
task~$A$ to write the shared data before modifying it. Task~$C$ waits
for all other tasks to enter the barrier before reading the
chunk. Note that some builtin functions are provided by the S-DSM
client API to get information about the topology (for example
\verb!clientGetClientNr! returns the number of S-DSM clients
instantiated in this run).

Some implicit synchronization also occurs when accessing shared data,
depending on the consistency protocol: READ, WRITE, READWRITE, PUT,
GET, readSymbol and writeSymbol are synchronous primitives that freeze
the process until another process releases the data.

\subsection{Event programming}

Event programming such as publish-subscribe is a model in which
actions are triggered based on specific events. This is a convenient
model for distributed applications, specifically written as a
distributed automata. While it is possible for the user to write
dynamic, event-base applications from scratch, it requires additional
error-prone code to manage events. This usually motivates the use of
dedicated event frameworks and, in this work, the integration of
publish-subscribe mechanisms directly within the
S-DSM~\cite{DBLP:conf/europar/Cudennec18}.

The publish-subscribe paradigm is defined by a set of mutable objects
(publishers) and a set of subscribers. There is a many-to-many
relationship between publishers and subscribers. Each time the mutable
object is changed, it publishes the information to all its
subscribers. The information can be a simple notification, an update
or the complete data. In this extension of the S-DSM API, chunks are
considered as mutable publishing objects. The distributed metadata
management for chunk coherence is extended on the S-DSM servers with
publish-subscribe metadata management. We consider the three listings
in Figure~\ref{fig:sdsm_publish_subscribe_listing}.

\begin{figure*}
  \centering
  \begin{lstlisting}
  void main_publisher() {
    mychunk = MALLOC(chunkid, size);
  /* wait for subscriber to subscribe to the chunk */
    rendezvous(RDV_ID);
    WRITE(mychunk);
     foo(mychunk);
    RELEASE(mychunk);
  }
\end{lstlisting}

\begin{lstlisting}
  void main_subscriber() {
    mychunk = LOOKUP(chunkid);
  /* subscribe to the chunk with given user handler */
    SUBSCRIBE(mychunk, subscriber_handler, params);
    /* we are ready for publish notifications */
    wakeup(RDV_ID);
  }
\end{lstlisting}

\begin{lstlisting}
  void subscriber_handler(chunk, params) {
    WRITE(chunk);
     foo(chunk);
    RELEASE(chunk);
/** unsubscribe to the chunk,
 *  this handler wont be call again */
    UNSUBSCRIBE(chunk);
/** afterwards, all publish notifications are discarded,
 * including the RELEASE in this function **/
  }
\end{lstlisting}

  \caption{Pseudo-code for publish-subscribe programming.}
  \label{fig:sdsm_publish_subscribe_listing}
\end{figure*}

The first listing implements the publisher role. This code only makes
use of regular S-DSM primitives. The publish-subscribe API is used by
subscribers, as presented by the second listing. The subscribe
primitive registers a user handler (a pointer to a local function) and
some user parameters to a given chunk. Each time the chunk is modified
-from anywhere in the S-DSM- this handler is called on the subscribing
task. Finally, a handler function example is given in the third
listing. Within the function it is possible to access shared data,
subscribe to other chunks and unsubscribe to any chunk. The same
handler function can be used to subscribe different chunks.

A user task is defined by a mandatory \verb!main! user function and
several optional \emph{handler} functions. The S-DSM runtime
bootstraps on the main function. At the end of this function, it falls
back to the builtin S-DSM client \emph{loop} function that waits for
incoming events such as publish notifications. If there are messages
postponed in the event pending list, then they are locally
replayed. If the task has no active chunk subscriptions, nor postponed
messages in the pending list, then it effectively terminates.

It is also possible to use signals: it works the same way as the
publish-subscribe mechanism except signals are standalone
synchronization objects not attached to chunks.

\section{Writing or adapting an application to the S-DSM}
\label{sec:application}

\begin{figure*}
  \centering
  \begin{lstlisting}
    static void
    prod(_SAT_Bootstrap_t * bootstrap) {
      /* user code for the producer role */
    }

    static void
    cons(_SAT_Bootstrap_t * bootstrap) {
      /* user code for the consumer role */
    }
    
    int
    main(int argc, char ** argv) {

      _SAT_Roles_t roles[3] = {NULL, prod, cons};

      _SAT_BOOTSTRAP(roles, NULL, argc, argv);
 
      return 0;
    }
  \end{lstlisting}
  \caption{Bootstrapping the S-DSM from the user code.}
  \label{fig:sdsm_bootstrap_user_code}
\end{figure*}

\begin{figure*}
  \centering
  \begin{lstlisting}
<?xml version="1.0"?>
<!DOCTYPE SAT>
<SAT xmlns:xsi="http://www.w3.org/2001/XMLSchema-instance">
 <topologies>
   <topology id="0" role="0">
     <memory capacity="0" />  
     <clients><intlist>1 2</intlist>
     </clients>
   </topology>
   <topology id="1" role="1">
     <memory capacity="0" />  
     <servers><intlist>0</intlist>
     </servers>
   </topology>
   <topology id="2" role="2">
     <memory capacity="0" />  
     <servers><intlist>0</intlist>
     </servers>
   </topology>
 </topologies>
</SAT>
  \end{lstlisting}
  \caption{Topology description.}
  \label{fig:sdsm_topology_description}
\end{figure*}

\begin{figure*}
  \centering
  \begin{lstlisting}
    mpirun -np 3 prodcons/prodcons --input bmp/lena_256x256.bmp --output out.bmp --topology prodcons/1server.xml

    mpirun -np 2 videostream/videostream --topology videostream/topology_pubsub_sched.xml : -np 1 videostream/videostreampthread : -np 1 videostream/videostreamopenmp : -np 1 videostream/videostreamopencl : -np 2 videostream/videostreamopencv camera window
  \end{lstlisting}
  \caption{Application command line.}
  \label{fig:sdsm_command_line}
\end{figure*}

The S-DSM is similar to a regular distributed or parallel application
that has been written as a set of processes and threads. Here are the
main steps to write or adapt an application.

\begin{description}

\item[Task model.] The user code is organized as a set of functions
  that implement different client roles. A role can be instantiated
  into several concurrent processes and is identified by a positive
  integer. By convention identifier~$0$ is the S-DSM server role and
  identifiers greater than~$0$ are user-defined client
  roles. Figure~\ref{fig:sdsm_bootstrap_user_code} shows two functions
  \verb!prod! and \verb!cons! implementing producer and consumer
  roles. User functions must comply to the same signature, taking a
  pointer to the S-DSM bootstrap structure and returning void. The
  user functions are registered in the main program into the roles
  structure. This structure is an array of pointers to functions
  indexed by the role identifier. Therefore, the first entry
  corresponds to the S-DSM server (it is set to NULL) and the
  following entries to the \verb!prod!  (role~$1$) and \verb!cons!
  (role~$2$) user-defined roles. This roles structure is given as a
  parameter of the bootstrapping primitive to run the corresponding
  code.

\item[Bootstrap.] The bootstrapping primitive is used to prepare the
  S-DSM environment and synchronize with other participating nodes. It
  must be called by all processes. One of the first step is to
  initialize the message passing runtime and retrieve the unique
  identifier of the task (the MPI rank in this implementation). By
  convention process~$0$ is a S-DSM server, also used as a \verb!seed!
  to bootstrap the distributed system. All processes contact the
  \verb!seed! to retrieve information about the neighborhood and the
  role they have been assigned. S-DSM clients get their corresponding
  server and servers get their S-DSM clients. All processes then enter
  a global distributed barrier before starting the server code or the
  user code corresponding to their role, hence starting the
  application from the user point of view. When the user code returns,
  the client sends a notification to its server and the bootstrapping
  primitive terminates on that client. Once a server receives
  termination notifications from all its clients it sends in turn a
  notification to the \verb!seed!. When all servers have notified the
  \verb!seed!  for termination, including itself, a directive is sent
  to all servers to shutdown the S-DSM. This bootstrap and termination
  protocol ensures that the S-DSM is up and running during the
  application life-time and that all requests are fulfilled.

\item[Topology.] The topology defines the number of instances
  (processes) per role and the connections between instances of S-DSM
  clients and instances of S-DSM
  servers. Figure~\ref{fig:sdsm_topology_description} gives an example
  of a simple topology made of one mandatory server (process~$0$
  playing role~$0$) to which are connected two clients (process~$1$
  playing role~$1$ and process~$2$ playing role~$2$). S-DSM topology
  is independent from the application and the same description can be
  applied to several applications. For example, using the topology
  described in Figure~\ref{fig:sdsm_topology_description} with the
  application written in Figure~\ref{fig:sdsm_bootstrap_user_code}
  will instantiate one S-DSM server process, one S-DSM client process
  running the \verb!prod! code and one client process running the
  \verb!cons! code according to the roles structure
  \verb!{NULL, prod, cons}!. Topology is written into an XML file that
  is parsed, serialized and partially transmitted to other processes
  by the \verb!seed! at bootstrap.

\item[Command line.] In this implementation MPI is used as the
  communication backend. The command line follows the MPI requirements
  with a mix of MPI parameters, S-DSM parameters and
  application-specific parameters. Figure~\ref{fig:sdsm_command_line}
  gives two examples of the S-DSM command line. The first command
  spawns $3$~MPI processes using the same binary
  (\verb!prodcons!). Parameter \verb!--topology! indicates to the
  \verb!seed! server where to find the topology description. The other
  parameters are left to the user code. If used with the topology
  given in Figure~\ref{fig:sdsm_topology_description}, MPI process
  with rank~$0$ will run the S-DSM server code and the ranks~$1$
  and~$2$ will respectively run \verb!prod! and \verb!cons!  user
  functions. The mapping of the MPI processes onto physical resources
  is described using regular MPI \verb!hostfile! and \verb!rankfile!
  files. The second command spawns $7$~MPI processes using different
  binaries. These binaries contain the same user function/role (a
  function that processes a frame in this example) implemented using
  different technologies to be mapped onto heterogeneous
  resources. MPI assigns rank to processes following the declaration
  order in the command line. In this example rank~$4$ will run
  binary~\verb!videostreamopencl! and must be mapped onto a resource
  that can execute OpenCL programs.
  
\end{description}

Writing the topology (application sizing), deciding the mapping
(resource allocation) and building the command line (technology
selection) reveals to be a complex task with severe performance issues
if the chosen configuration does not fit to the application and the
execution platform. As a follow-up of our experiments onto different
heterogeneous platforms, some work is conducted at CEA to
automatically explore some configurations in order to build a Pareto
front. From this Pareto representation it is possible to select a
particular configuration depending on some given constraints such as
the number of resources, the expected performance, latency, bandwidth
and the power consumption.

\subsection{Logging and profiling}

\begin{figure*}
  \centering
  \begin{lstlisting}
    /* bootstrapping processes */
    /* 0 = server (seed), 1 = server, 2,3 = clients */
    0 Bootstrap SAT MPI houghsat id 0 np 7 on host localhost with pid 3819
    2 Bootstrap SAT MPI houghsat id 2 np 7 on host localhost with pid 3821
    1 Bootstrap SAT MPI houghsat id 1 np 7 on host localhost with pid 3820
    ...
    
    /* parsing topology */
    0 Topology (7 topologies) loaded from file topology/2servers.xml
    /* start the seed server */
    0 Starting code (role 0)
    /* serialize topology on demand */
    0 Received message type 1 (request_topology) from 1
    0 Received message type 1 (request_topology) from 2
    ...
    
    /* allow all other processes to start */
    1 Starting code (role 0)
    2 Starting code (role 1)
    3 Starting code (role 2)
    ...
  \end{lstlisting}
  \caption{S-DSM debug example for bootstrapping, selected and
    commented parts. Timestamps have been removed for the sake of
    clarity.}
  \label{fig:sdsm_debug_bootstrap}
\end{figure*}

\begin{figure*}
  \centering
  \begin{lstlisting}    
    /* client 2 allocates chunk @1000 and asks for the write lock */
    2 malloc baseid 1000 size 256
    2 [Home-Based MESI] write chunk 1000@0 local state 3 (invalid)

    /* server 1 receives write request */
    1 Received message type 4 (consistency) from 2
    1 [Home-Based MESI] Server switch request 0 (client_req_write) from 2

    /* write request transferred to the home-node of chunk @1000 */
    0 Received message type 4 (consistency) from 1
    0 [Home-Based MESI] Server switch request 1 (server_req_write) from 1
    0 retrieve chunk 1000 version 0 entry version 0
    ...
    
    /* client 2 releases chunk @1000 */
    2 [Home-Based MESI] release chunk 1000@0 version 0 local state 1 (exclusive)
    /* client 2 uploads modified chunk to its server */
    1 Received message type 3 (data_ctrl) from 2
    1 update local chunk 1000@0 version 0 with version 1

    /* chunk release multi-hops to the home-node */
    1 Received message type 4 (consistency) from 2
    1 [Home-Based MESI] Server switch request 3 (client_req_release) from 2
    1 [Home-Based MESI] client req release for chunk 1000 version 1
    0 Received message type 4 (consistency) from 1
    0 [Home-Based MESI] Server switch request 4 (server_req_release) from 1
    0 RELEASE state 1 client 2 chunk 1000 version 1 metadata version 0    
  \end{lstlisting}
  \caption{S-DSM debug example for write section, selected and
    commented parts. Timestamps have been removed for the sake of
    clarity.}
  \label{fig:sdsm_debug_write_release}
\end{figure*}

\begin{figure*}
  \begin{center}
    \leavevmode
    \subfloat[Communication heatmap (in MB)]{
      \label{fig:sdsm_stats_heatmap}
      \includegraphics[width=0.5\textwidth]{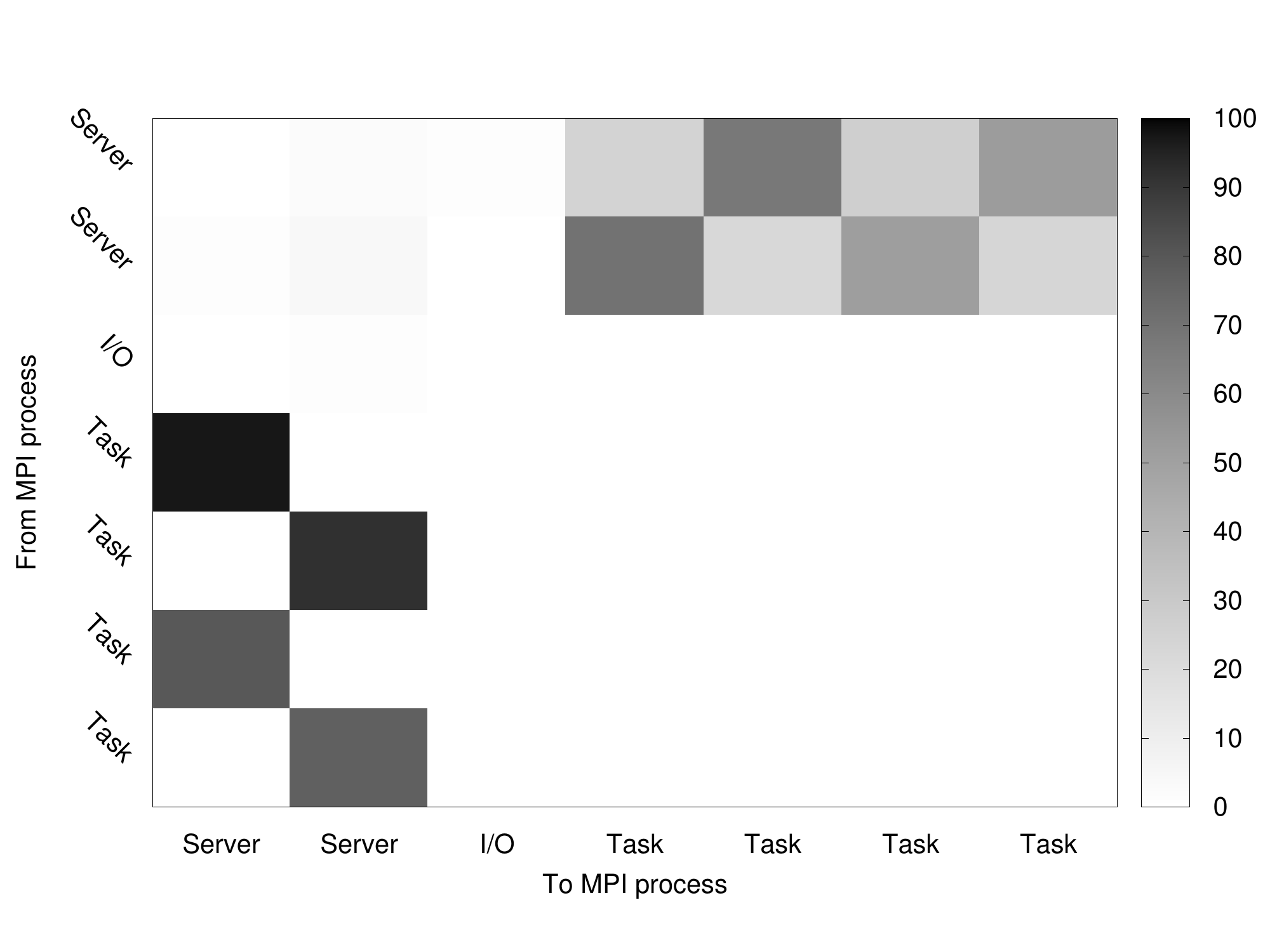}}
    \subfloat[Time decomposition and S-DSM overhead]{
      \label{fig:sdsm_stats_overhead}
      \includegraphics[width=0.5\textwidth]{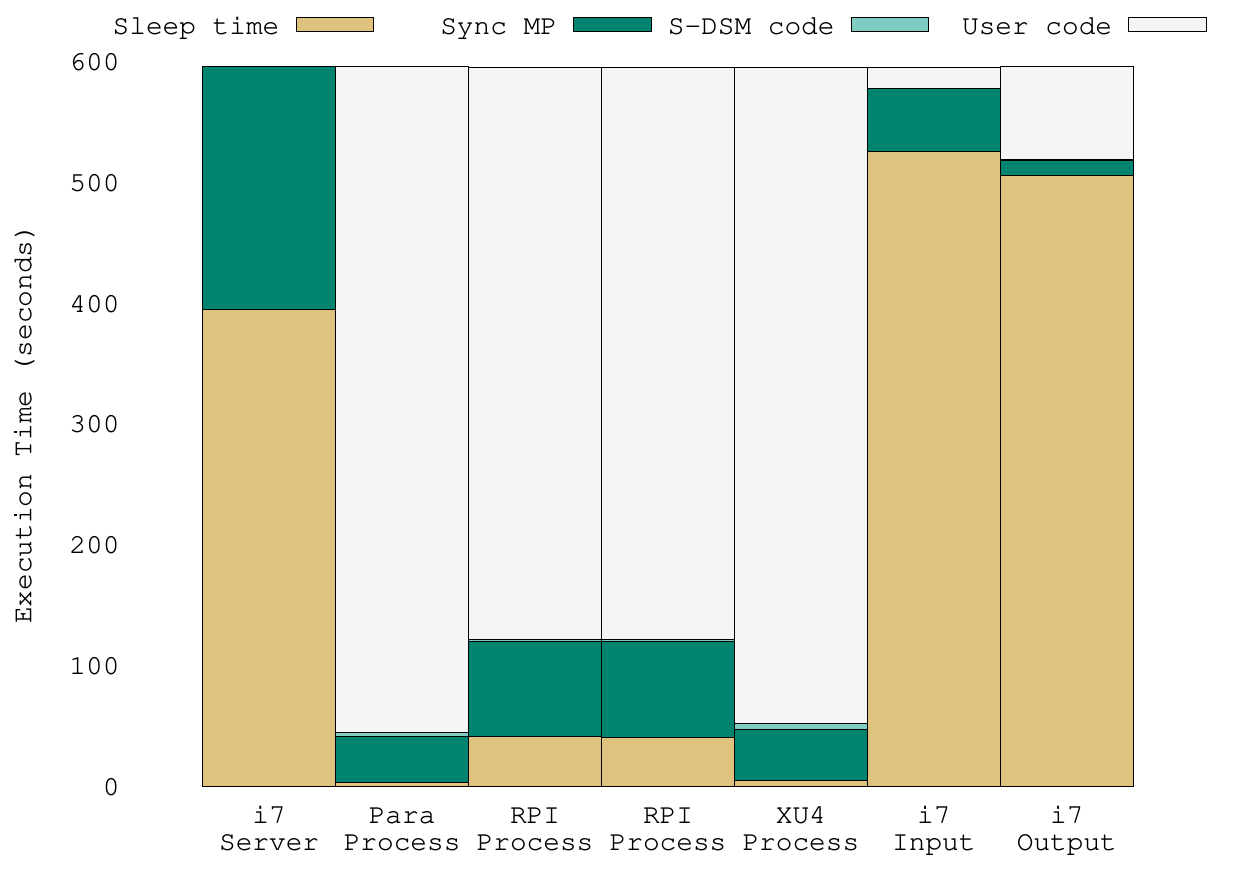}}
    \\
    \subfloat[Chunk allocation on a specific process]{
      \label{fig:sdsm_stats_alloc}
      \includegraphics[width=0.5\textwidth]{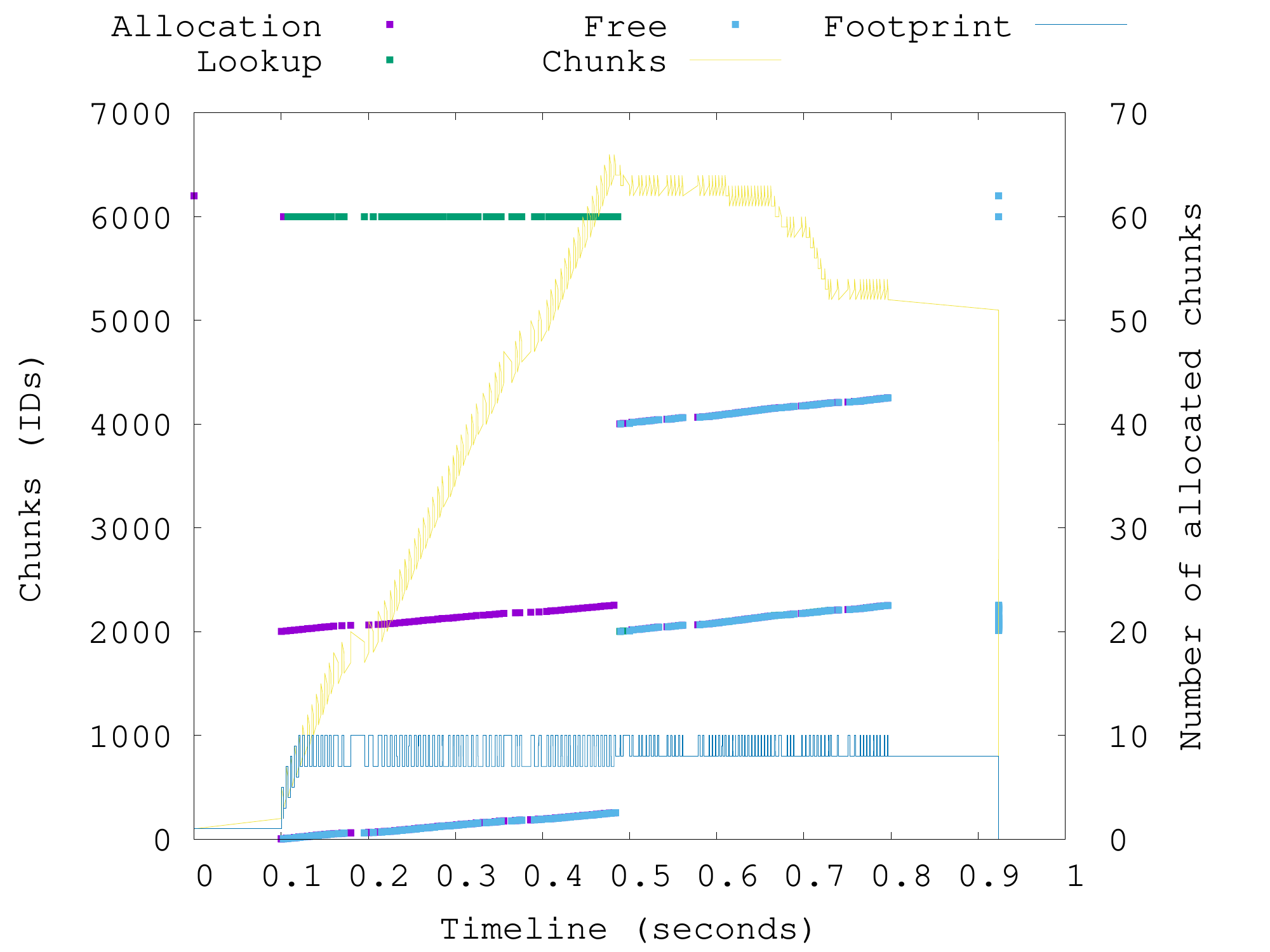}}
    \subfloat[Chunk access on a specific process]{
      \label{fig:sdsm_stats_consistency}
      \includegraphics[width=0.5\textwidth]{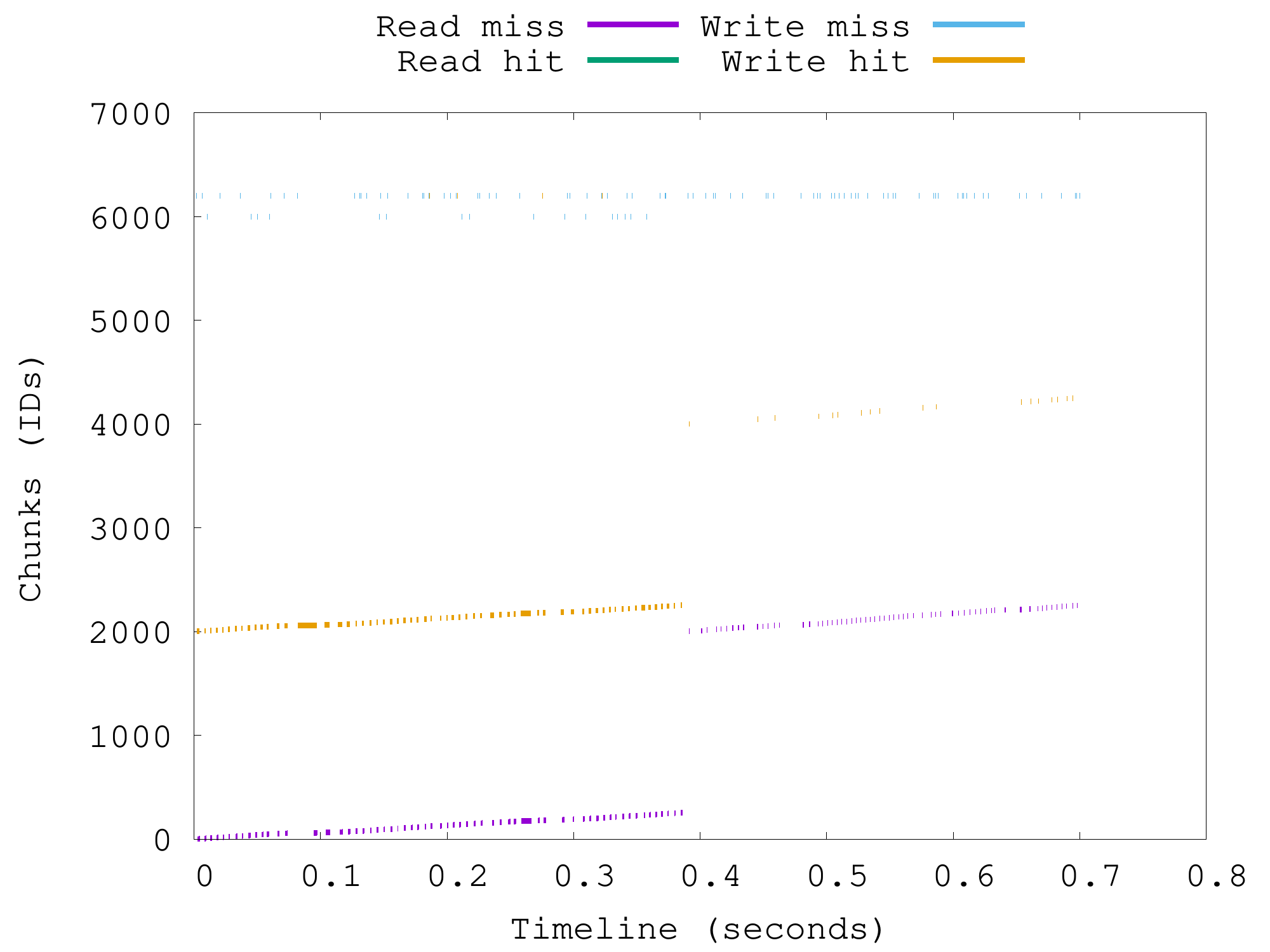}}

    \caption{S-DSM statistics.}
    \label{fig:sdsm_stats}
  \end{center}
\end{figure*}

Logging S-DSM events is used to debug and optimize applications. This
S-DSM can output two information streams. The \verb!debug! stream
similar to a verbose option for which all processes write events into
the standard output (\verb!stdout!). In this mode, the MPI runtime
conveniently aggregates all standard outputs into a single stream
located on the original master node. As for any distributed systems,
log entries are only ordered following causal dependencies making the
intricacy of events
nondeterministic~\cite{DBLP:journals/cacm/Lamport78}. Figures~\ref{fig:sdsm_debug_bootstrap}
and~\ref{fig:sdsm_debug_write_release} give two examples of a
\verb!debug! stream for bootstrapping and opening a write
section. This stream generates multiple system calls to \verb!stdout!,
additional network traffic to aggregates the log onto the master node,
as well as a potentially large file if redirected with the \verb!&>!
  operator. As a result the performance of the S-DSM runtime can be
  severely affected and the analysis of the access patterns to shared
  data might lead to conclusions that do not apply when running
  without debug. A more verbose option can also be used to dump the
  content of messages to the log.

The \verb!statistics! stream logs events that are related to
performance and tuning of the application. On each process, internal
events are continuously stored into the local physical memory and
dumped to local files when the S-DSM terminates. Unlike the debug
stream, this strategy does not significantly affect performances and
can be used to analyze memory access patterns. However, it may
allocate an important amount of local physical memory and this can
lead to performance issues in case of a large number of S-DSM events
or a long run. It is thereafter possible to use a script to analyze
and generate figures. Some examples in Figure~\ref{fig:sdsm_stats}
include:

\begin{description}

\item[Communication heatmap~\ref{fig:sdsm_stats_heatmap}] represents
  the cumulative amount of messages in MB sent between processes. This
  map is divided into $4$~parts. Server-to-server communications are
  quite light: the home-based MESI consistency protocol mainly
  generates short control messages. Server-to-client communications
  mainly consists in data transfers to update clients on access
  request. Client-to-server communications are used to upload the new
  version of data after an exclusive access. It also shows the chosen
  topology between clients and servers, as a client can only contact
  its associated server. Finally, client-to-client communications are
  not allowed in this consistency protocol implementation.

\item[Time decomposition~\ref{fig:sdsm_stats_overhead}] shows the time
  spent for each process in different parts of the user and S-DSM
  code. The \verb!user code! corresponds to the time spent in the
  application code, excluding S-DSM calls. The \verb!S-DSM code! time
  corresponds to the local data management, excluding all consistency
  protocol communications. \verb!Sync MP! is the time spent in the
  message passing send and receive primitives, excluding the sleep
  time. Finally, \verb!Sleep! is the time spent in sleep mode while
  waiting for an incoming message. This implementation in the S-DSM
  runtime is based on micro-sleeping with a loop call to
  \verb!clock_nanosleep! using adaptable sleep times. It has been
  introduced to limit the energy consumption that occurs when polling
  for new messages, as it is designed in most of the MPI
  runtimes. More information about the micro-sleeping mechanism can be
  found in this article~\cite{doi:10.1002/cpe.5960}. From the time
  decomposition, \verb!Sleep! and \verb!user code! times can be
  interpreted as an efficient use of resources, while \verb!Sync MP!
  and \verb!S-DSM code! times can be considered as overhead.

\item[Chunk allocation~\ref{fig:sdsm_stats_alloc}] gives, for a
  particular process, some events related to chunk allocation during
  the execution time. The left ordinate represents the S-DSM logical
  address space and the right ordinate represents the number of
  allocated chunks on the process. Events include the allocation, the
  lookup (retrieving a previously allocated chunk) and free (locally
  removing the data, not necessarily the chunk metadata). The
  \verb!Chunks! curve shows the number of chunks locally known by the
  process and the \verb!Footprint! corresponds to the number of chunks
  with a valid (allocated) pointer to the data. In this example, a
  limit has been set to $10$~chunks after which other chunks are
  locally evicted using a LRU policy.

\item[Chunk access~\ref{fig:sdsm_stats_consistency}] is quite similar
  to the chunk allocation figure and shows the miss and hit events for
  both read and write access types for a particular process. Events
  represent the entire consistency scope, starting from the acquire
  primitive (read, write, readwrite), including the communications, to
  the release primitive. Therefore, a long segment means that the
  chunk has been locked for this entire time, with possible contention
  if other processes are waiting in the pending list of this chunk.

\end{description}

\subsection{The video stream processing application}

\begin{figure*}
\centering
  \includegraphics[width=\textwidth]{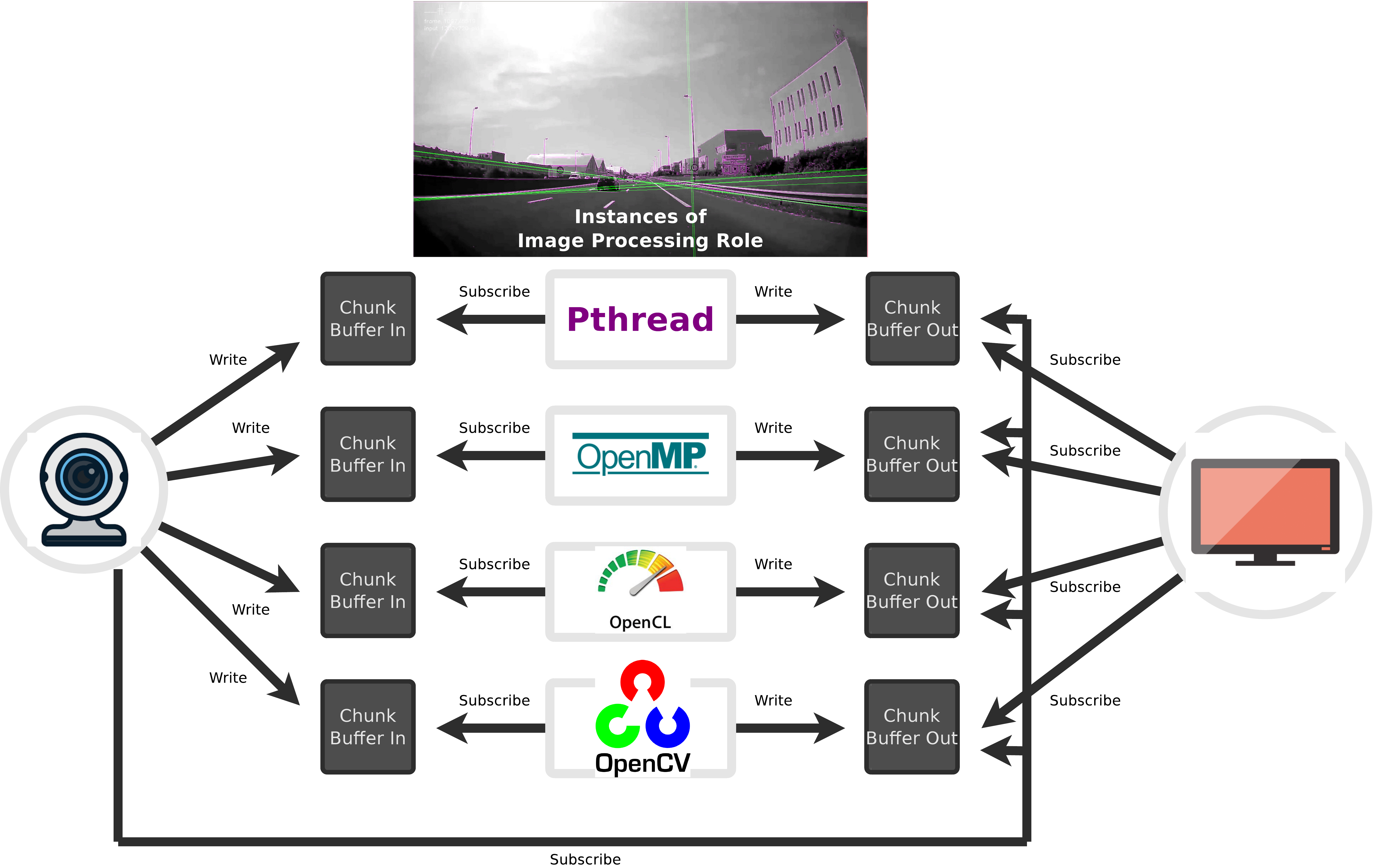}
  \caption{Application for line detection in a video stream.}
  \label{fig:sdsm_application_videostream}
\end{figure*}

The video stream processing application has been designed to run on
heterogeneous platforms using the S-DSM programming API. This
application is made of $3$~roles: an \verb!input! role decodes a video
into raw frames from a file or a webcam. It dispatches the frame to
one of the \verb!process! roles that calculates the resulting image
and an \verb!output! role encodes back the processed frames into a
file or a live monitoring window.

A \verb!process! role applies an edge detection followed by a line
detection (Hough transform) on the input frame. This role can be
instantiated several times. Edge detection is implemented using a
$3$x$3$~convolution stencil. While the convolution complexity is
constant, the Hough transform complexity is data-dependent: the
complexity differs from one frame to another. Above a detection
threshold, a pixel is represented as a sinusoid in the intermediate
transformed representation. In this intermediate representation, above
a second detection threshold, a pixel is represented as a line in the
final output image. Both transform operations require the use of
double-precision sinus and cosinus functions, which is quite demanding
in terms of computational power. The \verb!process! role has been
written in different technologies: sequential C, Pthread, OpenMP,
OpenCL and OpenCV, using the builtin OpenCV functions. This allows to
choose a suitable implementation when mapping instances of this role
onto heterogeneous resources.

The design of the video processing application is very close to a
dataflow application as shown in
Figure~\ref{fig:sdsm_application_videostream}, in which a set of tasks
communicate using explicit channels. In this application, channels are
implemented thanks to shared buffers, which is a classical approach in
NUMA machines. The novelty is that the S-DSM allows to keep this
shared buffer implementation among distributed systems.

For each \verb!process! task, one input buffer and one output buffer
are allocated in the S-DSM to store the input and processed
frames. There is no explicit synchronization in the user code to
manage data between tasks: this synchronization comes by design with a
mix of exclusive write accesses and publish-subscribe
notifications. Whenever a raw frame has been decoded from the input
stream, the \verb!input! task writes into an available input
buffer. This notifies the corresponding processing task that a new
frame is ready and calculates the resulting image into its output
buffer. The \verb!input! and \verb!output!  tasks are notified of this
write: the \verb!output! task encodes the result while the
\verb!input!  task flags the corresponding input buffer as available
to receive a new raw frame.

This implements de-facto a dynamic scheduler based on eager policy.
This is quite convenient if the processing tasks are deployed on
heterogeneous resources with a strong variability in computing time.

The videostream application has been used to do experiments on
heterogeneous testbeds~\cite{DBLP:conf/europar/Cudennec18} and to build a M2DC
demonstrator at Teratec and ISC High Performance Frankfurt 2018.

\section{Conclusion}
\label{sec:conclusion}

Designing and building a distributed shared memory requires to
implement sophisticated mechanisms including local memory management
and distributed algorithms. The combination of these mechanisms makes
the global system complex to tune and debug. However this S-DSM is
implemented using basic concepts and well documented algorithms: chunk
management is a common approach in peer-to-peer systems since their
massive deployment in the early 2000, and most of the distributed
synchronization objects have been largely studied forty years ago with
contributions from Lamport and Raynal. Despite its simplicity and
sometimes the use of naive concepts, this S-DSM, as for other modern
S-DSM, is able to compete with message-passing systems. This allows to
write applications using shared memory over distributed architectures
while not paying the price of a significant computing time
overhead. In some configurations, we are even able to observe better
performance due to a wise management of data and the inner
introduction of pipeline parallelism when storing data on intermediate
S-DSM servers. However, a significant part of computing performance
comes from the fine tuning of the middleware, including making smart
choices for the configuration of the run. This highlights the
importance of proposing decision-making tools at compile time such as
operational research algorithm (including AI) to configure the
deployment.

\section*{Acknowledgments}
This work has received funding from the European Union’s Horizon 2020
research and innovation programme under grant agreement No 688201.

\bibliographystyle{plain}
\bibliography{report}

\end{document}